\documentclass[pra,twocolumn,superscriptaddress,amsmath,amssymb,floatfix,noshowpacs]{revtex4}
\usepackage{dcolumn}
\usepackage{bm}
\usepackage{graphicx,hyperref}
\usepackage{color}
\usepackage{soul}
\usepackage{amsmath}
\usepackage{tabularx}
\graphicspath{{fig/}}
\usepackage{amsmath,amssymb}
\usepackage{centernot}
\usepackage{stmaryrd}
\usepackage{natbib}

\begin{document}

\title{\textbf{
Heisenberg-Langevin approach to driven superradiance}}

\author{Ori Somech}
\thanks{Present Address: Centre for Quantum Dynamics, Griffith University, Brisbane Queensland 4111, Australia}
\affiliation{Department of Chemical \& Biological Physics, Weizmann Institute of Science, Rehovot 7610001, Israel}
\author{Yoav Shimshi}
\affiliation{Department of Chemical \& Biological Physics, Weizmann Institute of Science, Rehovot 7610001, Israel}
\author{Ephraim Shahmoon}
\affiliation{Department of Chemical \& Biological Physics, Weizmann Institute of Science, Rehovot 7610001, Israel}
\date{\today}

\begin{abstract}
We present an analytical approach for the study of driven Dicke superradiance based on a Heisenberg-Langevin formulation. We calculate the steady-state fluctuations of both the atomic-spin and the light-field operators. While the atoms become entangled below a critical drive, exhibiting spin squeezing, we show that the radiated light is in a classical-like coherent state whose amplitude and spectrum are identical to those of the incident driving field. Therefore, the nonlinear atomic system scatters light as a linear classical scatterer. Our results are consistent with the recent theory of coherently radiating spin states. The presented Heisenberg-Langevin approach should be simple to generalize for treating superradiance beyond the permutation-symmetric Dicke model.
\end{abstract}

\maketitle

\section{\label{sec:Introduction} Introduction}
Superradiance describes the cooperative radiation of an ensemble of quantum emitters into common photonic modes. A conceptually simple case that captures the essence of cooperative radiation is that of Dicke superradiance, where all the constituents of an ensemble of two-level atoms are coupled to the common photonic modes in an identical manner, thus forming an effective ``collective spin" dipole \cite{Dicke, mandel_wolf_1995,GH}. Superradiance was observed both in atoms \cite{HAR,TOMs1,TOMs2,FLD,BRWsr} and artificial emitters \cite{MAJ} and plays a role in various quantum phenomena and technologies, ranging from phase transitions \cite{KES,EMAN} to narrowband superradiant lasers \cite{HAK,HOL,TOM1,MOL1}.

The situation wherein the atoms are additionally driven by a resonant laser can be studied by a driven-dissipative master equation of the Dicke model. Mean-field theory yields a second order phase transition of the steady-state atomic population, or ``magnetization", as a function of the drive \cite{DRUMMOND1978160,DRUMMOND1980,CAR,LAW,LAR,BAR}. Spin squeezing was recently found in steady state by a numerical solution of the master equation \cite{Alejandro,yelin,BAR,REYt} with a supporting analytical result obtained in \cite{yelin}. For the radiated light, intensity correlations $g^{(2)}$ were calculated and found to exhibit bunching correlations above the phase-transition point, but no correlations below it \cite{CAR}. More recently, it was found that the appearance of so-called coherently radiating spin states (CRSS) as the steady-state of driven Dicke superradiance underlies these results \cite{CRSS}.

Here we present a simple analytical approach for driven superradiance based on Heisenberg-Langevin (HL) equations. While this HL approach is in principle equivalent to the master equation used previously, the HL equations are natural for the direct analytical treatment of both spin and field fluctuations via their operator-form solution. In particular, we account for spin and field fluctuations around the mean field using the Holstein-Primakoff approximation. For the spin fluctuations the operator-valued solutions are in a Bogoliubov transformation form implying quantum correlations, as verified by the subsequent analytical calculation of spin squeezing. For the field operator, we find that the fluctuations are proportional to the vacuum field, thus proving that the radiated field below the phase-transition point is in a coherent state. We also calculate the two-time correlation of the field, finding that the spectrum is delta-peaked at the incident-drive frequency. Surprisingly, the light is thus scattered from the many-atom system as if the latter is a linear system, although the atomic system is highly nonlinear, as evident by its phase transition. We discuss the consistency and relation of these results with the predictions of CRSS theory \cite{CRSS}. 

The paper is organized as follows. In Sec. II we derive the HL equations of the driven Dicke model, focusing on a relevant cavity-scheme realization. After recalling the mean-field solution in Sec. III, we treat spin fluctuations and squeezing in Sec. IV. Sec. V is devoted to the analysis of the radiated light. Finally, our conclusions are presented in Sec. VI.

\section{Model}
We begin with the derivation of the HL equations of motion that describe the driven Dicke model, considering a system of atoms in a damped cavity as realized in typical experiments \cite{HAR,TOMs1,TOMs2,MAJ}. Realizations of Dicke physics exist also in other systems wherein many atoms are coupled to a common photon bath, e.g. in waveguide QED \cite{Alejandro} or even in an elongated atomic ensemble in free space \cite{GH,BRWsr}; however, the cavity case considered here is conceptually the most straightforward one as it directly emphasizes a single common photonic mode.

\begin{figure}
    \centering
    \includegraphics[width=\columnwidth]{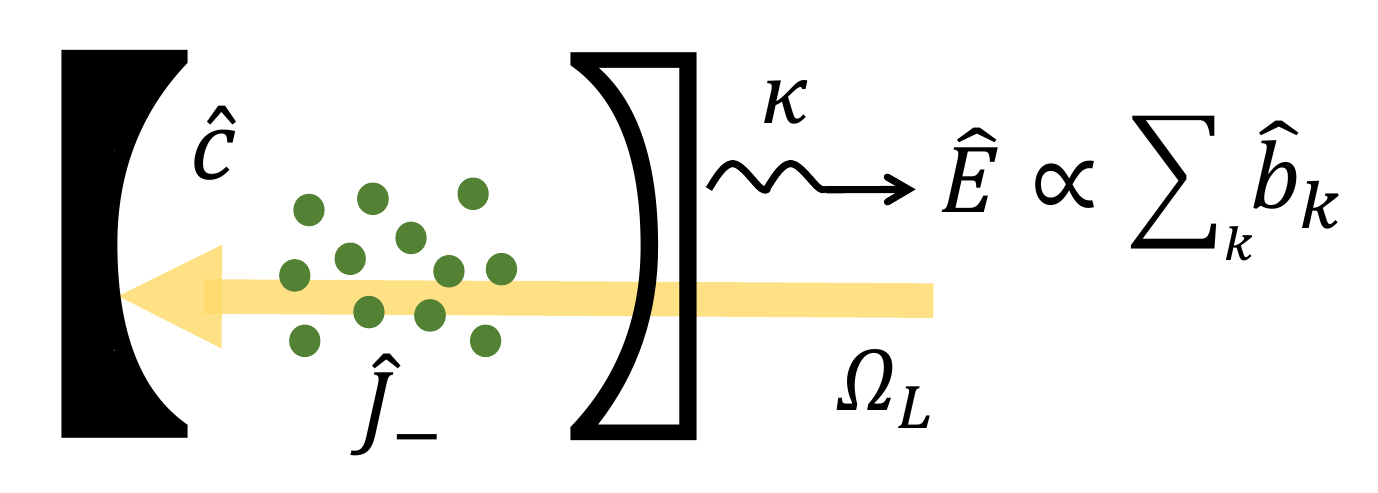}
    \caption{Cavity realization of driven superradiance. An atomic ensemble is trapped inside a cavity, wherein all atoms (green dots) are identically coupled to a cavity mode (lowering operator $\hat{c}$) and hence described by a collective-spin dipole (lowering operator $\hat{J}_-$). The cavity field is damped through its mirrors at rate $\kappa$ to the outside propagating modes $\hat{b}_k$, which form the radiated field $\hat{E}$, and is driven by a laser with Rabi-field amplitude $\Omega_L$. Here a single-sided cavity scheme is presented, with one out-coupling mirror (right-hand side).}
    \label{fig:physicalrealization}
\end{figure}

\subsection{System and Hamiltonian}
We consider the system displayed in Fig. 1: $N$ two-level atoms are trapped inside an optical cavity driven by external laser light thorough the cavity mirrors. The atomic positions are such that all atoms are identically coupled to the cavity mode (i.e. well within the cavity mode waist and at longitudinal positions that are multiples of cavity wavelength apart). The Hamiltonian of the atoms and the cavity is given by
\begin{equation}
\hat{H}_{S}=\hbar\omega_a\hat{J}_z+\hbar\omega_{c}\hat{c}^{\dagger}\hat{c}+\hbar\left[\hat{c}^\dagger\left(g^{\ast}\hat{J}_-+\Omega_{L}e^{-i\omega_{L}t}\right)+\text{h.c.}\right].
\label{Hs}
\end{equation}
Here $\hat{c}$ is the boson lowering operator of the cavity mode of frequency $\omega_c$, whereas $\hat{J}_{\alpha}=(1/2)\sum_{n=1}^N\hat{\sigma}_n^{\alpha}$ ($\alpha\in\{x,y,z\}$) are the collective-spin operators of the atomic ensemble with $\hat{\sigma}_n^{\alpha}$ being the Pauli operator of a two-level atom $n\in\{1,...,N\}$ with resonant frequency $\omega_a$. The cavity is driven via its mirrors by an external laser of frequency $\omega_L$ and amplitude $\Omega_L$, and is coupled to the atoms via the dipole coupling $g$ identical to all atoms, where $\hat{J}_-=\hat{J}_x-i\hat{J}_y=\sum_{n=1}^N\hat{\sigma}_n^{-}=\hat{J}_+^{\dag}$ is the collective-spin lowering operator of the atoms and $\hat{\sigma}_n^{-}=(\hat{\sigma}_n^{+})^{\dag}$ the Pauli lowering operator of atom $n$.

In addition, the cavity mode is coupled through its mirrors to a 1D continuum of propagating photon modes characterized by the wavenumber $k$ and corresponding boson modes $\hat{b}_k$ and frequencies $vk$ ($v$ being the speed of light). The Hamiltonians describing this 1D photon reservoir and its coupling to the system are given by, respectively (here, for one-sided cavity, Fig. 1),
\begin{eqnarray}
\hat{H}_{R}&=&\sum_{k>0}\hbar v k\hat{b}_{k}^{\dagger}\hat{b}_{k},
\nonumber\\
\hat{H}_{SR}&=&\hbar\sum_{k>0}\left(\eta\hat{b}_{k}^{\dagger}\hat{c}+\text{h.c.}\right), \quad \eta\equiv\sqrt{\frac{v}{L}\kappa},
\end{eqnarray}
where the coupling constant $\eta$ is taken $k$-independent (consistent with the Markov approximation) and $L$ is the quantization length of the 1D continuum. The total Hamiltonian is given by $\hat{H}=\hat{H}_{S}+\hat{H}_{R}+\hat{H}_{SR}$. We note that we neglect here the direct spontaneous emission from atoms to photon modes in transverse directions outside the cavity. For a dilute ensemble this is an individual-atom process that is typically much slower than the relevant Dicke dynamics discussed here.

\subsection{Heisenberg-Langevin equations}
We begin with eliminating the reservoir modes $\hat{b}_k$ by inserting the solution of their Heisenberg equations into the equation for $\hat{c}$, obtaining within the usual Markov approximation \cite{SCU}
\begin{equation}
\dot{\tilde{c}}=\left(i\delta_c-\frac{\kappa}{2}\right)\tilde{c}-ig^{\ast}\tilde{J}_- -i\Omega_L+\hat{E}_0(t), \quad \delta_c=\omega_L-\omega_c.
\label{cEOM}
\end{equation}
Here the system operators are already written in a rotated frame, $\tilde{c}=\hat{c}e^{i\omega_L t}$ and  $\tilde{J}_-=\hat{J}_- e^{i\omega_L t}$, whereas the Langevin, vacuum noise of the reservoir is given by $\hat{E}_0(t)=-i\sum_k\eta^{\ast}e^{-i(vk-\omega_L)t}\hat{b}_k(0)$, satisfying (assuming an initial vacuum state)
\begin{equation}
\langle \hat{E}_0(t)\hat{E}_0^{\dag}(t')\rangle=\kappa \delta(t-t').
\label{E0}
\end{equation}

Next, we eliminate the cavity mode by assuming that its damping rate $\kappa$ to the 1D continuum is much faster than the typical time scale of variations in $\tilde{J}_-$, i.e. $\kappa\gg |\dot{\tilde{J}}_-/ \tilde{J}_-|$. Within this coarse-grained dynamical picture and for times $t$ much longer than $1/\kappa$, the elimination of $\tilde{c}$ is equivalent to setting $\dot{\tilde{c}}=0$ in Eq. (\ref{cEOM}) and inserting the solution for $\tilde{c}$ into the Heisenberg equations for atomic variables such as $\tilde{J}_-$ and $\hat{J}_z$. Finally, we obtain (denoting $\tilde{J}_{\mp}\rightarrow \hat{J}_{\mp}$ for simplicity)
\begin{eqnarray}
\dot{\hat{J}}_-&=&i\delta\hat{J}_-+\left(\gamma-i2\Delta\right) \hat{J}_z\hat{J}_- -i2\hat{J}_z\left[\Omega+\hat{f}(t)\right],
\nonumber\\
\dot{\hat{J}}_{z}&=&-\gamma\hat{J}_+\hat{J}_-+i\hat{J}_+\left[\Omega+\hat{f}(t)\right]-i\left[\Omega^{*}+\hat{f}^{\dag}(t)\right]\hat{J}_-,
\nonumber\\
\label{HL}
\end{eqnarray}
with the laser-atom detuning $\delta=\omega_L-\omega_a$, the coefficients
\begin{equation}
\gamma=\frac{|g|^2\kappa}{\delta_c^2+(\kappa/2)^2},
\quad
\Delta=\frac{-|g|^2\delta_c}{\delta_c^2+(\kappa/2)^2},
\quad
\Omega=\frac{-2g \Omega_L}{2\delta_2+i\kappa},
\label{par}
\end{equation}
and the effective Langevin, input-vacuum noise (filtered by the cavity), $\hat{f}(t)\approx [2g/(\kappa-i2\delta_c)]\hat{E}_0(t)$, satisfying
\begin{equation}
\langle \hat{f}(t)\hat{f}^{\dag}(t')\rangle=\gamma \delta(t-t').
\label{f}
\end{equation}
Equations (\ref{HL}) form the HL equations of the driven Dicke model, with an effective emission rate $\gamma$ of an atom to the outside modes via the cavity, and an effective laser drive with Rabi frequency $\Omega$. The collective shift $\Delta$ describes the resonant dipole-dipole interactions between pairs of atoms \cite{LEH}, corresponding to an effective Hamiltonian $\hat{H}_{\text{dd}}=-\hbar\sum_{n}\sum_m\Delta_{nm}\hat{\sigma}_{n}^{+}\hat{\sigma}_{m}^-$. Here the dipole-dipole kernel $\Delta_{nm}=\Delta$ is uniform for all atom pairs $n$ and $m$ since all atoms are coupled identically to the mediating cavity photon mode. In treatments of superradiance in free space, such coherent dipole-dipole effects are often ignored in free-space \cite{GH} whereas they vanish in a waveguide QED superradiance scheme \cite{Alejandro}. In the cavity setting, they exist however if one allows for laser-cavity detuning $\delta_c$
as seen in Eq. (\ref{par}) for $\Delta$ and noted in Refs. \cite{BAR,REYt}.

We note that while this specific derivation was performed starting from the damped-cavity model, equivalent HL equations (\ref{HL}) can be derived by considering other models of photon continua to which all atoms are identically coupled. Here the cavity mode effectively becomes a continuum due to its fast damping rate $\kappa$.

\subsection{Equivalent master equation}
The HL equations (\ref{HL}) are equivalent to the following master equation for the density matrix of the atoms,
\begin{eqnarray}
\frac{d\hat{\rho}}{d t}&=&-\frac{i}{\hbar}\left[\hat{H}_{\text {eff }},\hat{\rho}\right]+\gamma \left[\hat{J}_-\hat{\rho} \hat{J}_+-\frac{1}{2}\left(\hat{J}_+\hat{J}_-\hat{\rho}+\hat{\rho}\hat{J}_+\hat{J}_-\right)\right],
\nonumber\\
\hat{H}_{\text {eff}}&=&-\hbar\Delta\hat{J}_+\hat{J}_--\hbar\left(\Omega \hat{J}_+ +\Omega^{\ast}\hat{J}_-\right).
\label{ME}
\end{eqnarray}
Here we have already assumed that the laser drive is resonant with the atoms, $\delta=\omega_L-\omega_a=0$. This master equation with $\Delta=0$ is a typical starting point for the analysis of driven Dicke superradiance presented in previous works \cite{DRUMMOND1978160,DRUMMOND1980,CAR,LAW,LAR,Alejandro,yelin,RABk}, whereas the additional dipole-dipole term $\Delta$ is considered in Refs. \cite{BAR,REYt}. Here instead we will use the HL formulation of Eq. (\ref{HL}), in order to derive analytical results for fluctuations and correlations of atomic and photonic degrees of freedom. We will use the master equation as a numerical verification of the one-time correlation functions of the atoms. Since the total spin $\hat{J}_x^2+\hat{J}_y^2+\hat{J}_z^2=j(j+1)$ is conserved under the dynamics of Eqs.(\ref{HL}) and (\ref{ME}), the initial state sets the SU(2) spin representation $j$. Assuming an initial ground state for the $N$ atoms, we have $j=N/2$ and the Hilbert space that spans Eqs. (\ref{ME}) is of size $2j+1=N+1$ and can be easily solved numerically for reasonable $N$.

\section{Mean-field solution}
We begin with the mean-field solution of the model in steady state. To obtain the mean-field equations, we take the average over the HL equations (\ref{HL}), such that the Langevin vacuum-noise terms vanish, and perform the factorization of operator products $\langle \hat{J}_\alpha \hat{J}_\beta \rangle\approx\langle \hat{J}_\alpha \rangle \langle \hat{J}_\beta \rangle$ (with $\alpha,\beta\in\{x,y,z\}$).
This factorization is justified for $N\rightarrow \infty$ under the mean-field assumption that fluctuations of observables are much smaller than their mean. It is important to note that such a factorization does not mean that there are no correlations between the atoms that comprise the collective spin $\hat{J}_{\alpha}$ \cite{DRUMMOND1978160}: in fact, we see below that the atoms are entangled \cite{Alejandro,yelin,BAR,REYt}. Considering the conservation of the total spin $\hat{J}_x^2+\hat{J}_y^2+\hat{J}_z^2=j(j+1)$ with $j=N/2\gg 1$ and taking a resonant drive $\delta=0$, the solution to the mean-field equations becomes (see also \cite{BAR}),
\begin{eqnarray}
\langle \hat{J}_z\rangle=-\frac{N}{2}\sqrt{1-\frac{|\Omega|^2}{\Omega_c^2}},
\quad
\langle \hat{J}_-\rangle=-\frac{\Omega}{\Delta+i\gamma/2},
\label{MF}
\end{eqnarray}
with the critical driving field defined by
\begin{eqnarray}
\Omega_c=\Omega_c(\Delta)=\frac{N}{4}\sqrt{\gamma^2+4\Delta^2}.
\label{Oc}
\end{eqnarray}
The steady-state population inversion (or ``magnetization") $\langle \hat{J}_z\rangle$ thus exhibits a second order phase transition as a function of the drive $\Omega$, where it vanishes at the critical value $\Omega_c$. The latter increases with the strength of the dipole-dipole shift $\Delta$ as seen in Eq. (\ref{Oc}). For $|\Omega|>\Omega_c$ there exist oscillatory solutions of the mean-field equations \cite{DRUMMOND1978160} which nevertheless appear to decay to zero at long time scales, upon the consideration of the full quantum problem \cite{BAR}. Figure 2 displays $\langle \hat{J}_z\rangle$ obtained by the exact numerical solutions of the master equation (\ref{ME}) for $N=50$ and different values of $\Delta$. Very good agreement with the mean-field expression (\ref{MF}) is exhibited when $|\Omega|$ is not too close to the critical point $\Omega_c$. In particular, calculations with different values of $\Delta$ all collapse to the same curve when $\Omega$ is scaled to the corresponding $\Omega_c(\Delta)$ from Eq. (\ref{Oc}). Disagreement between mean-field and numerical results is observed around $\Omega_c$ due to the fact that the mean value of $\langle \hat{J}_{z}\rangle$  near $\Omega_c$ becomes increasingly small while fluctuations grow, in contradiction to the mean-field assumption. The second order transition predicted by the mean-field solution in the thermodynamic limit $N\rightarrow \infty$ then becomes smoother at finite $N$.
\begin{figure}
    \centering
    \includegraphics[width=\columnwidth]{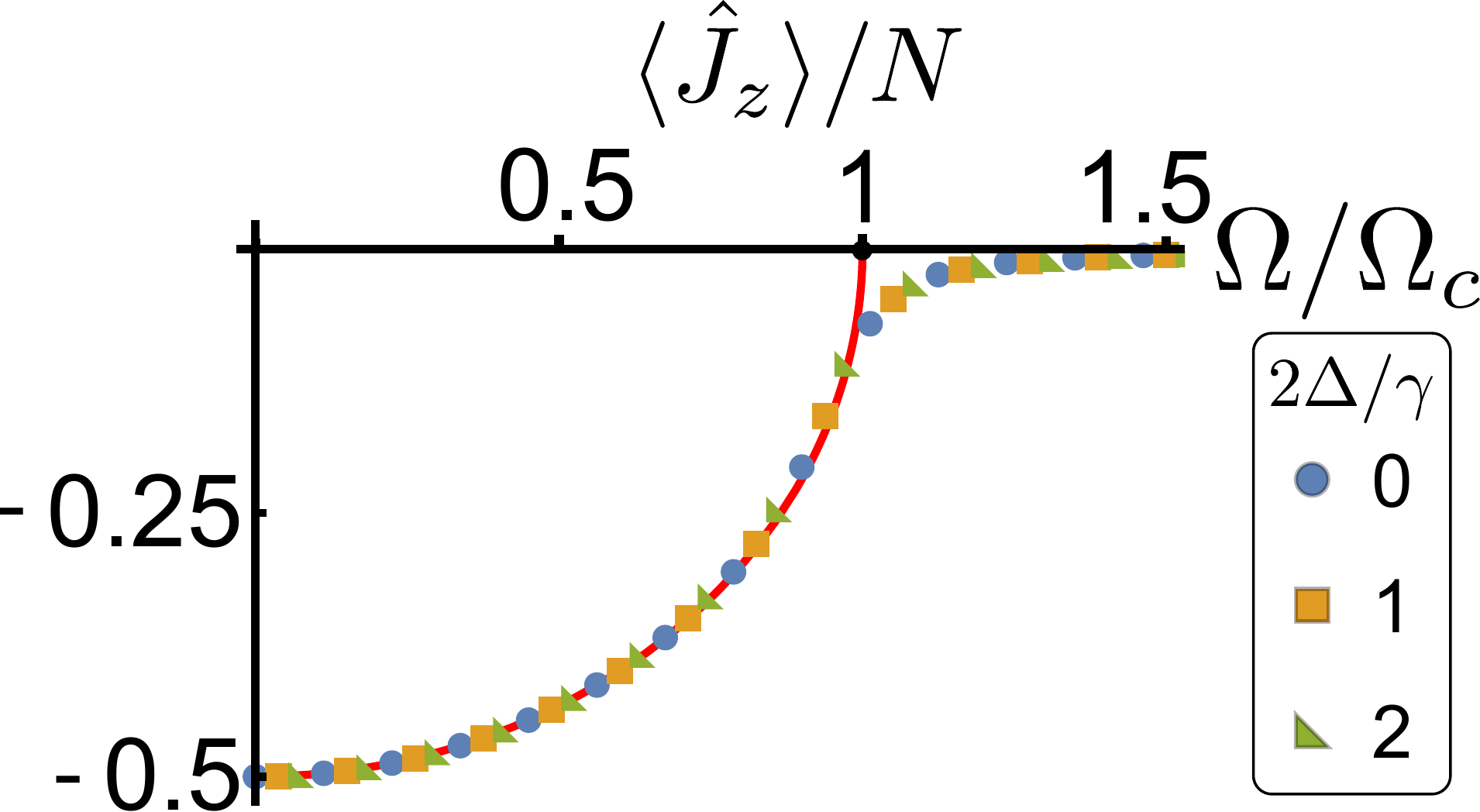}
    \caption{Population inversion $\langle\hat{J}_z\rangle$ of the collective atomic system as function of the driving field $\Omega$. Results obtained by the numerical solution of Eq. (\ref{ME}) with $N=50$ atoms and for different values of the dipole-dipole shift $2\Delta/\gamma=0,1,2$ all collapse to the same curve when $\Omega$ is scaled the critical field $\Omega_c (\Delta)$ from Eq. (\ref{Oc}). The red line represent the analytical mean-field solution from Eq.~(\ref{MF}), exhibiting a second order phase transition. The exact numerical solutions agree with the mean-field result until they diverge away a bit before the transition point due to the finite value of $N$.}
    \label{fig:Jz}
\end{figure}

The mean-field solution (\ref{MF}) can be also written as a mean of the spin vector $\hat{\mathbf{J}}=(\hat{J}_x,\hat{J}_y,\hat{J}_z)$ in a Bloch sphere,
\begin{eqnarray}
\langle \hat{\mathbf{J}}\rangle=
\left(
        \begin{array}{c}
          \langle \hat{J}_x\rangle \\
          \langle \hat{J}_y\rangle \\
          \langle \hat{J}_z\rangle \\
        \end{array}
\right)
=-\frac{N}{2}
\left(
        \begin{array}{c}
          \sin\theta\cos\phi \\
          \sin\theta\sin\phi \\
          \cos\theta\\
        \end{array}
\right),
\label{MSV}
\end{eqnarray}
with the angles in spherical coordinates given by
\begin{eqnarray}
\sin\theta=\frac{|\Omega|}{\Omega_c}, \quad \phi=\mathrm{arg}(\Delta+i\gamma/2)-\mathrm{arg}(\Omega).
\label{theta}
\end{eqnarray}
For later purposes, it is instructive to introduce a rotated coordinate system at which the mean spin vector is directed to the south pole of the Bloch sphere and hence appears as a ground state in this rotated system. Spin operators in the rotated system, described by the vector $\hat{\mathbf{J}}'=(\hat{J}'_x,\hat{J}'_y,\hat{J}'_z)$ are related to the original spin operators $\hat{\mathbf{J}}=(\hat{J}_x,\hat{J}_y,\hat{J}_z)$ via the rotation matrix $\mathcal{R}$ as
\begin{eqnarray}
&&\hat{\mathbf{J}}'=\mathcal{R}^{-1}\hat{\mathbf{J}},
\quad \langle \hat{\mathbf{J}}'\rangle=-\frac{N}{2}
\left(
        \begin{array}{c}
          0 \\
          0 \\
          1 \\
        \end{array}
\right),
\nonumber\\
&&\mathcal{R}=\left(
  \begin{array}{ccc}
    \cos\theta\cos\phi & -\sin\phi & \sin\theta\cos\phi \\
    \cos\theta\sin\phi & \cos\phi & \sin\theta\sin\phi \\
    -\sin\theta & 0 & \cos\theta \\
  \end{array}
\right).
\label{R}
\end{eqnarray}
As required, in the rotated system the mean spin vector $\langle \hat{\mathbf{J}}'\rangle$ points to the south pole, defining the $-z'$ axis as the mean spin direction.

\section{Spin fluctuations and squeezing}
We now turn to the analysis of small fluctuations of spin variables around the mean-field solution (Sec. IV A). This will allow us to estimate atomic correlations such as spin squeezing (Sec. IV B), and later on also the fluctuations in the scattered field (Sec. V).

\subsection{Collective spin fluctuations in the Holstein-Primakoff approximation}
We recall that within its representation in the rotated system (\ref{R}), the mean spin vector $\hat{\mathbf{J}}'=(\hat{J}'_x,\hat{J}'_y,\hat{J}'_z)$ is directed towards the axis $z'$ and vanishes along the $x',y'$ axes.
In order to analyze fluctuations around this mean, we first define the spin lowering operator in the rotated basis, $\hat{J'}_-=\hat{J}'_{x}-i\hat{J}'_{y}$, and re-write the HL equations (\ref{HL}) in terms of the rotated-spin operators $\hat{J}'_-,\hat{J}'_{z}$ using the transformation $\mathcal{R}$ from (\ref{R}). As in the original basis, the HL in the rotated basis are also nonlinear in their relevant variables, $\hat{J}'_-,\hat{J}'_{z}$; however, the linearization of the equations for small fluctuations around the mean field is simpler in this rotated basis. To this end, we use the Holstein-Primakoff transformation, which is an exact representation of SU(2) spin operators (here of spin $j=N/2$) in terms of a bosonic operator $\hat{a}$ (satisfying $[\hat{a},\hat{a}^{\dag}]=1$) \cite{ASA},
\begin{equation}
\hat{J}'_-=\sqrt{N-\hat{a}^{\dagger}\hat{a}} \ \hat{a}, \quad \hat{J}'_{z}=\hat{a}^{\dagger}\hat{a}-\frac{N}{2}.
\label{HP}
\end{equation}
We see that the limit $\hat{a}\rightarrow 0$ is that of the mean-field solution (\ref{MSV}), so that the vacuum of $\hat{a}$ is the mean field and $\hat{a}$ describes fluctuations on top of it. In line with the mean-field assumption, we consider small fluctuations, $|\hat{a}|\sim O(1)\ll \sqrt{N}$, and expand the nonlinear HL equation for $\hat{J}'_-$ to leading orders in the small parameter $1/\sqrt{N}$. This is achieved by the approximation
\begin{equation}
\hat{J}'_-\approx \sqrt{N} \ \hat{a}, \quad \hat{J}'_{z}\approx-\frac{N}{2},
\label{HPa}
\end{equation}
and the subsequent linearization of the HL equation to first orders of $\hat{a}$ and the noise $\hat{f}$. Finally, we obtain the HL equation for the spin fluctuations $\hat{a}$,
\begin{eqnarray}
\dot{\hat{a}}&=&-\left(N\frac{\gamma}{2}\cos\theta-iN\Delta\frac{1+\cos^2\theta}{2}\right)\hat{a}-i N \Delta\frac{\sin^2{\theta}}{2}\hat{a}^{\dag}
\nonumber\\
&+&i\sqrt{N}\left[\frac{1+\cos\theta}{2}e^{i\phi}\hat{f}(t)-\frac{1-\cos\theta}{2}e^{-i\phi}\hat{f}^{\dag}(t)\right].
\label{HLa}
\end{eqnarray}
This yields coupled linear equations for $\hat{a}$ and $\hat{a}^{\dag}$ whose solution in the steady state for times $t\gg (N \gamma \cos\theta/2)^{-1}$ is
\begin{eqnarray}
\hat{a}(t)&=&\sqrt{N}\left[\frac{1+\cos\theta}{2}e^{i\phi}\hat{B}(t)+\frac{1-\cos\theta}{2}e^{-i\phi}\hat{B}^{\dag}(t)\right],
\nonumber\\
\hat{B}(t)&=&i\int_0^t dt' e^{-N\cos\theta\left[\frac{\gamma}{2}-i\Delta\right](t-t')}\hat{f}(t').
\label{a}
\end{eqnarray}
This operator-form solution, along with the correlation function (\ref{f}) of the Langevin vacuum-noise $\hat{f}(t)$, now allows to evaluate correlations of the collective spin. In fact, even without performing specific calculations, the operator solution itself is already quite insightful. We see that the lowering operator of the spin fluctuation $\hat{a}$ exhibits a Bogoliubov transformation form: it is a linear combination of the integrated vacuum noise lowering operator $\hat{B}$ and its conjugate $\hat{B}^{\dag}$, with corresponding Bogoliubov coefficients proportional to $1+\cos\theta$ and $1-\cos\theta$, respectively. Non-trivial, correlated fluctuations occur whenever $\hat{a}$ contains the conjugate component $\hat{B}^{\dag}$ (and not only $\hat{B}$), requiring a non-vanishing coefficient $1-\cos\theta$. Therefore, quantum correlations are expected to grow with the driving field $|\Omega|/\Omega_c=\sin\theta>0$ , as seen explicitly below.

\subsection{Spin squeezing}
A particulary relevant characterization of collective-spin fluctuations is provided by the spin squeezing parameter \cite{spinsqueezingreview,KitaUeda}. Spin squeezing quantifies fluctuations of the spin vector perpendicular to its mean direction, and is linked to the sensitivity of quantum-enhanced metrology with collections of spins \cite{spinsqueezingparameter1,spinsqueezingparameter2,QSr} and their underlying pairwise entanglement \cite{LEW,SOR}. Within the rotated spin representation from (\ref{R}), where the mean is directed to $-z'$, the spin squeezing parameter is given by \cite{spinsqueezingparameter1,spinsqueezingparameter2,spinsqueezingreview}
\begin{equation}
\xi^{2}=\mathrm{min}_{\varphi} \frac{\mathrm{Var}[\hat{J}'_{\varphi}]N}{|\langle\hat{J}'_z\rangle|^2},
\quad
\hat{J}'_{\varphi}=\cos\varphi \hat{J}'_x+\sin\varphi\hat{J}'_y,
\label{xi}
\end{equation}
i.e. it is proportional to the minimal variance of the fluctuations along the $x'y'$ plane. Spin squeezing exists for $\xi^{2}<1$, implying that the collective-spin has improved phase sensitivity to rotations compared to the standard quantum limit $\xi^2=1$ of an uncorrelated coherent spin state.

Within our mean-field and small-fluctuations assumption, we use $|\langle\hat{J}'_z\rangle| \approx N/2$ and the bosonic approximation (\ref{HPa}) for $\hat{J}'_{\mp}=\hat{J}'_x\mp i\hat{J}'_y$, to obtain the spin squeezing parameter in terms of the bosonic operators $\hat{a}$,
\begin{equation}
\xi^{2}=1+2\langle \hat{a}^{\dagger}\hat{a}\rangle-2|\langle\hat{a}^{2}\rangle|.
\label{xia}
\end{equation}
Using the solution for $\hat{a}$, Eq. (\ref{a}), and the Langevin, vacuum-noise correlation function (\ref{f}), we then find
\begin{equation}
\left|\langle a^{2}\rangle \right|=\frac{1-\cos^2\theta}{4\cos\theta},
\quad \langle a^{\dagger}a\rangle =\frac{(1-\cos\theta)^2}{4\cos\theta},
\label{aa}
\end{equation}
so that the spin squeezing parameter, Eq. (\ref{xia}), becomes
\begin{equation}
\xi^{2}=\cos\theta=\sqrt{1-\frac{|\Omega|^2}{\Omega_{c}^2(\Delta)}}.
\label{xis}
\end{equation}
We observe that the spin squeezing is determined by the ratio between the driving field and the critical field, $|\Omega|/\Omega_c$. It depends on the dipole-dipole interaction $\Delta$ through the critical field $\Omega_c(\Delta)$ from Eq. (\ref{Oc}). This generalizes the analytical result of Ref. \cite{yelin}, obtained for the case $\Delta=0$ using a master-equation approach. When the drive is weak $|\Omega|/\Omega_c\rightarrow 0$, no spin squeezing exists, $\xi^2=1$, since the system is in a coherent spin state wherein all atoms are in the ground state.
As the drive increases, population in the atoms is created, such that collective emission is possible, building entanglement and spin-squeezing correlations between the atoms, $\xi^2<1$. At the critical point $|\Omega|/\Omega_{c}=1$ the spin-squeezing parameter vanishes: this result is valid only at the limit $N\rightarrow\infty$ where it does not contradict the Heisenberg limit $\xi^{2}\geqslant1/N$ \cite{spinsqueezingreview}. For finite $N$, our mean-field assumption of small fluctuations breaks down as we approach the critical point, where fluctuations become increasingly large (e.g. $\mathrm{max}_{\varphi} \mathrm{Var}[\hat{J}'_{\varphi}]\propto 1/\cos\theta$ diverges near the critical point).

It is instructive to compare the analytical result (\ref{xis}) to that obtained by an exact numerical solution of the master equation for a finite $N$, as explained above. In Fig. 3 we observe excellent agreement between the analytical and numerical solutions up to a driving field somewhat below the critical point $|\Omega|<\Omega_{c}(\Delta)$, above which the two solutions diverge away. As in Fig. 2, the dependence on $\Delta$ is captured by plotting the numerical solutions for different values of $\Delta$, which all collapse to the curve as a function of the driving field $\Omega$ (e.g. taken real) scaled to the corresponding critical field $\Omega_c(\Delta)$, as anticipated analytically in Eq. (\ref{xis}). 
We observe that the exact solution obtains its optimal (minimal) value for the squeezing $\xi^2$ close to the point where it begins to diverge away from the analytical result. Therefore, this optimal value for $\xi^2$ should improve (become smaller) with increasing $N$ \cite{yelin}.
The scaling of the optimal $\xi^2$, being a finite-size effect, cannot be accounted for by the above mean-field based results (valid for $N\rightarrow\infty$). This scaling can be obtained analytically using CRSS theory, yielding $\xi^2\sim N^{-1/3}$ \cite{CRSS}.
\begin{figure}
    \centering
    \includegraphics[width=\columnwidth]{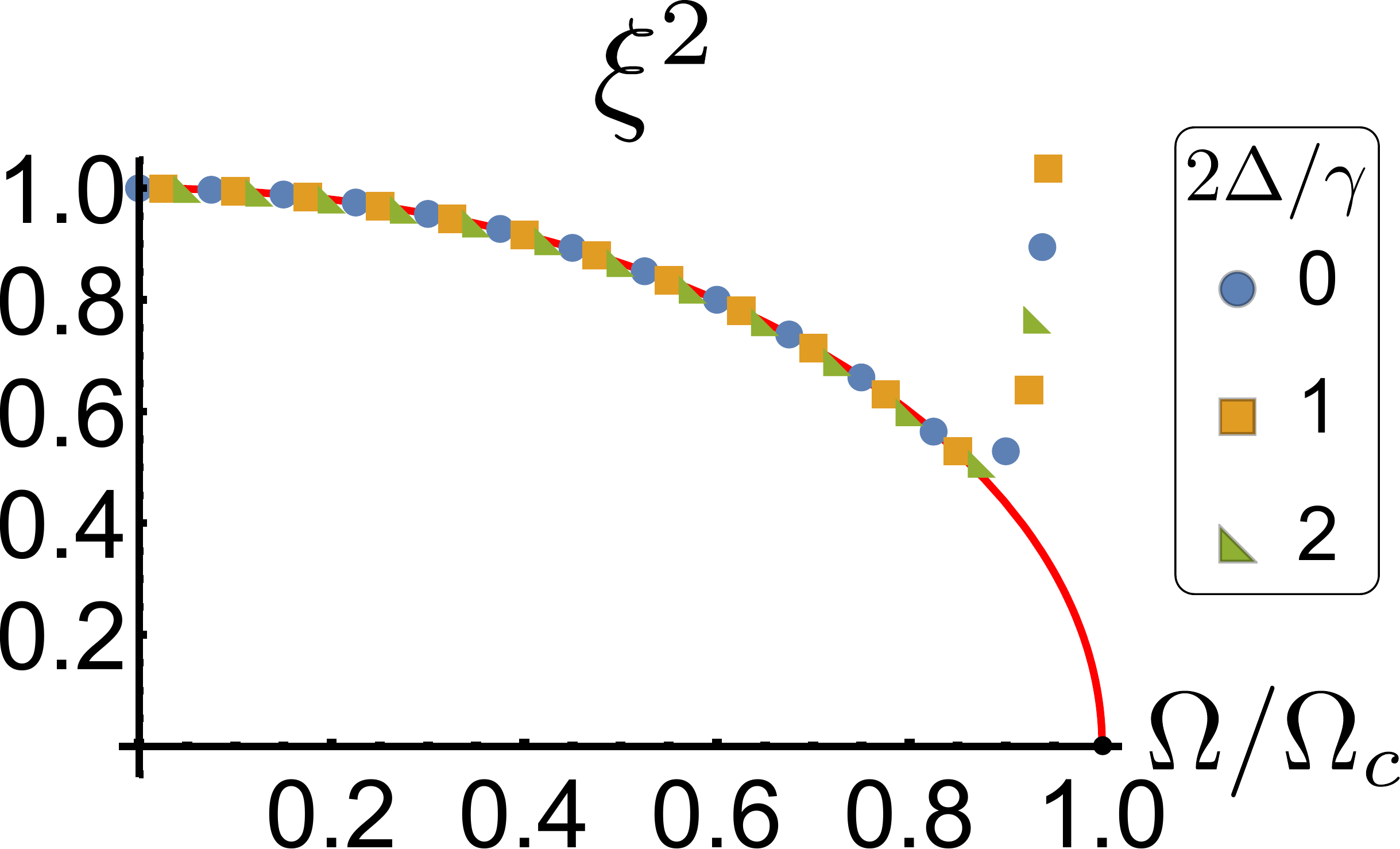}
    \caption{Spin squeezing $\xi^{2}$ as function of the driving field $\Omega$. Results obtained by the numerical solution of Eq. (\ref{ME}) with $N=50$ atoms and for different values of the dipole-dipole shift $2\Delta/\gamma=0,1,2$ all collapse to the same curve by when $\Omega$ is scaled the critical field $\Omega_c (\Delta)$ from Eq. (\ref{Oc}). The red line represent the analytical solution from Eq.~(\ref{xis}). The exact numerical solutions agree with the analytical result until they diverge away close to the transition point, where $\xi^{2}$ begins degrading (growing) with $\Omega$, see main text.}
    \label{fig:spinsqueezing}
\end{figure}

\section{Radiated Light}
So far we have treated the field degrees of freedom as a reservoir that generates driven-dissipative dynamics of the atoms. However, superradiance is essentially a scattering problem of an input coherent-state field off a collective dipole $\hat{J}_-$ formed by the atoms. As such, the total field exhibits the general form,
\begin{equation}
\hat{E}(t)=\hat{E}_{\mathrm{free}}(t)+G \hat{J}_-.
\label{E1}
\end{equation}
The first term is the freely propagating coherent-state field in the absence of atoms, consisting of an average field and vacuum fluctuations. It may include the influence of linear optical elements such as the cavity mirrors in the cavity realization of Fig. 1. The second term is the field component scattered by the atomic dipole $\hat{J}_-$, with a coupling coefficient $G$ (describing field propagation from the atoms to the detector). While the first term exhibits non-correlated coherent-state statistics of the input field, the second term may exhibit correlations generated by the nonlinearity of the atoms \cite{QNLOr}. In superradiance, the considered atomic system is clearly nonlinear, as we have already seen that the population inversion $\langle\hat{J}_z\rangle$ is a nonlinear function of the driving field $\Omega$, see Eq. (\ref{MF}). Nevertheless, we show in the following that, surprisingly, the scattered component of the field is also a coherent-state field, linear in the input field. This holds for any driving field $\Omega$ smaller than the critical field $\Omega_c$.

Although this result is valid for any realization of superradiance, we focus for concreteness on the cavity realization considered above. We define the total observable field as the field propagating out of the cavity (in the rotated frame $\omega_L$)
\begin{equation}
\hat{E}(t)=-i\sum_{k>0}\eta^{\ast}\hat{b}_k(t)e^{i\omega_L t}-i\Omega_L,
\label{E2}
\end{equation}
where $-i\Omega_L$ is the average component of the input coherent field. 
Using the same HL approach from Sec. II B, we solve for $\hat{E}(t)$ within the coarse-grained dynamics at $t\gg 1/\kappa$, obtaining 
Eq. (\ref{E1}) with (see Appendix),
\begin{eqnarray}
\hat{E}_{\mathrm{free}}(t)&=&\left(1+\chi\right)\left[\hat{E}_0(t)-i\Omega_L\right],
\quad
\chi=\frac{\kappa}{i\delta_c-\kappa/2},
\nonumber\\
G&=&-ig^{\ast}\chi.
\label{E3}
\end{eqnarray}
Here $\chi$ describes the linear response of the cavity to the input field $\hat{E}_0(t)-i\Omega_L$ (vacuum + coherent drive), which interferes with the input, yielding the factor $1+\chi$. Therefore, the atom-free field indeed has the form of a coherent-state field composed of vacuum + average components. In the following we will show that this turns out to be the case also for the total field.

\subsection{Average field}
Taking the average of Eq. (\ref{E1}), the vacuum term $\hat{E}_0$ does not contribute so that the free-field component from Eq. (\ref{E3}) gives $-i\Omega_L(1+\chi)$. For the scattered part we use $G=-ig^{\ast}\chi$ from Eq. (\ref{E3}) and $\langle \hat{J}_-\rangle$ from Eq. (\ref{MF}) obtaining $G\langle \hat{J}_-\rangle=i\chi \Omega_L$. The total average field then becomes,
\begin{equation}
\langle \hat{E}\rangle=-i\Omega_L,
\label{Ea}
\end{equation}
equal to the incident average field. So, the average radiated field in superradiance is linear in the incident-field amplitude even though the atomic system is nonlinear, as discussed above. 

\subsection{Field fluctuations}
The HL approach allows us to gain direct access to field operators which entail information on the quantum statistics of the field. We will use it here to show that the fluctuating part of the radiated field is proportional to vacuum fluctuations, thus proving that the radiated field is in a coherent state. We first do this by solving for the operators directly, without the need to infer the statistics from the calculation of correlations.

To this end, we focus on the scattered component of the field, $G\hat{J}_-$. Using the transformation (\ref{R}), we write $\hat{J}_-$ in terms of the rotated-system spin operators as,
\begin{eqnarray}
\hat{J}_-&=&e^{-i\phi}\left(\frac{\cos\theta+1}{2}\hat{J}'_- + \frac{\cos\theta-1}{2}\hat{J}'_+ + \sin\theta \hat{J}'_z \right)
\nonumber\\
&\approx& e^{-i\phi}\left(\frac{\cos\theta+1}{2}\sqrt{N}\hat{a} + \frac{\cos\theta-1}{2}\sqrt{N}\hat{a}^{\dag} - \sin\theta \frac{N}{2} \right).
\nonumber\\
\label{Jm1}
\end{eqnarray}
In the second line we have used the Holstein-Primakoff linearization, Eq. (\ref{HPa}).
Plugging in the solution for $\hat{a}$ from Eq. (\ref{a}), we then obtain for the fluctuating part of the field $\hat{E}$ from (\ref{E1})
\begin{eqnarray}
\hat{\mathcal{E}}(t)\equiv\hat{E}-\langle\hat{E}\rangle=\left(1+\chi\right)\hat{E}_0(t)+G N\cos\theta \hat{B}(t).
\nonumber\\
\label{dE}
\end{eqnarray}
The first term describes the vacuum fluctuations $\propto \hat{E}_0$ of the coherent free-field component from Eq. (\ref{E3}). The second term originates from the fluctuating part of the scattered field $G\hat{J}_-$ from Eq. (\ref{Jm1}) and is also essentially proportional to integrated vacuum fluctuations $\hat{E}_0$ [noting that $\hat{B}$ in Eq. (\ref{a}) is an integral of $\hat{f}\propto\hat{E}_0$].  This proves that the total radiated field is in a coherent state, comprised of vacuum fluctuations on top of a mean coherent amplitude.

\subsection{Light squeezing vs. spin squeezing}
Since the radiated field is a classical-like coherent state, it does not exhibit any quantum correlations. We now show this explicitly for the case of quantum squeezing correlations. Defining the quadrature operator of the radiated field, $\hat{X}_{\varphi}=e^{-i\varphi}\hat{E}+e^{i\varphi}\hat{E}^{\dag}$, the bosonic squeezing parameter of the field is given by
\begin{eqnarray}
\xi^2_E=\mathrm{min}_{\varphi}\frac{\mathrm{Var}[\hat{X}_{\varphi}]}{V_0}=1+\frac{2}{V_0}\left(\langle\hat{\mathcal{E}}^{\dag}\hat{\mathcal{E}}\rangle-|\langle\hat{\mathcal{E}}^2\rangle|\right),
\label{xiE}
\end{eqnarray}
with $V_0\equiv [\hat{E},\hat{E}^{\dag}]=[\hat{E}_0,\hat{E}^{\dag}_0]=\kappa \delta(t=0)$ being the vacuum-noise level. Squeezed quantum noise and correlations exist if the quadrature noise can become lower than that of the vacuum, i.e. for $\xi^2_E<1$. It is seen that this requires the existence of the phase-dependent correlator $\langle\hat{\mathcal{E}}^2\rangle$. Similarly, spin squeezing in Eq. (\ref{xia}) requires the existence of the phase-dependent correlator of spin fluctuations $\langle\hat{a}^2\rangle$. For either of these correlators to exist, the corresponding lowering operators $\hat{\mathcal{E}}$ and $\hat{a}$ then must contain a raising field-operator $\hat{B}^{\dag}$ (equivalently, $\hat{E}_0^{\dag}$) in addition to $\hat{B}$, since the average is performed over the initial vacuum state. While the Bogoliubov coefficient $1-\cos\theta$ in Eq. (\ref{a}) indeed guarantees that $\hat{a}$ contains $\hat{B}^{\dag}$ for any finite drive $|\Omega|/\Omega_c=\sin\theta<1$, this is not the case for the field fluctuations $\hat{\mathcal{E}}$: The transformation coefficients in Eq. (\ref{Jm1}) from $\hat{a},\hat{a}^{\dag}$ to $\hat{J}_-\sim \hat{\mathcal{E}}$, which also depend on $\cos\theta$, lead to an exact cancellation of the coefficient for $\hat{B}^{\dag}$ in $\hat{\mathcal{E}}$, as seen in Eq. (\ref{dE}). Therefore, for any drive strength $|\Omega|/\Omega_c=\sin\theta<1$, spin squeezing exists while light squeezing exactly cancels. This result is equivalent to the geometrical interpretation given by the so-called dipole-projected squeezing \cite{CRSS}.

\subsection{Spectrum}
Having access to the field operator 
$\hat{E}(t)$, the HL approach also allows to directly calculate two-time correlations and spectra. The spectrum of the radiated field in a steady state time $t$ is given as usual by the Fourier transform on the time-difference $\tau$ of the two-time correlation, $\langle\hat{E}^{\dag}(t)\hat{E}(t+\tau)\rangle$. Since this is a normal-ordered correlator, the fluctuating part of $\hat{E}$ in Eq. (\ref{dE}) drops, as it is proportional to the lowering operator $\hat{E}_0$. This trivially yields $\langle\hat{E}^{\dag}(t)\hat{E}(t+\tau)\rangle=|\langle\hat{E}\rangle|^2=|\Omega_L|^2$. The spectrum of the radiated field is then a single delta peak at the incident frequency $\omega_L$ (recalling we work in the laser-rotated frame). This again shows that the collective atomic dipole scatters light as a linear optical element even though the atomic population exhibits a strongly nonlinear dependence on the drive $\Omega_L$.

\section{Conclusions}
In this work we have presented a HL approach to driven Dicke superradiance in steady state. The analytical results for steady-state spin squeezing agree and generalize those obtained in Refs. \cite{Alejandro,yelin,BAR,REYt}. Furthermore, our finding that the radiated field is in a coherent state underlies previous results on uncorrelated photon statistics below the transition \cite{CAR}.
These HL-based results are consistent with the formation of a CRSS as described in \cite{CRSS}. The HL approach is thus complementary to the CRSS description of superradiance. On the one hand, it is based on the approximate analysis of small fluctuations around the mean field for $N\rightarrow \infty$, and did not yield the finite-size scalings with $N$ or the full atomic state as in CRSS. But on the other hand, it is simpler to generalize for treating superradiance beyond the permutation-symmetric Dicke case, e.g. by performing the Holstein-Primakoff approximation for each individual atom separately, while allowing for direct estimation of atom and field correlations.

\appendix*
\section{Output field}
Here we elaborate on the derivation of the general expression for the output field, Eqs. (\ref{E1}) and (\ref{E3}), in the one-sided cavity scheme of Fig. 1. We begin by defining the outside propagating field
\begin{eqnarray}
\hat{E}(x,t)=-i\sum_{k>0}\eta^{\ast}\hat{b}_k(t)e^{i k x}e^{i\omega_L t}-i\Omega_L e^{i\frac{\omega_L}{v}x}.
\label{A1}
\end{eqnarray}
Here $x$ is the propagation axis: in the one-sided scheme, $x=0$ denotes the position of the outcoupling mirror (right-hand side mirror in Fig. 1), so that $x<0$ denotes incoming left-propagating fields whereas $x>0$ denotes outgoing right-propagating fields. The radiated field from Eq. (\ref{E2}) is then defined by taking $x=0^+>0$. As in the derivation of the HL equations in Sec. II B, we first formally solve the Heisenberg equations for $\hat{b}_k(t)$, obtaining in the Markov approximation
\begin{eqnarray}
\hat{E}(x,t)&=&\hat{E}_0(x,t)-i\Omega_L e^{i\frac{\omega_L}{v}x}
\nonumber\\
&-&e^{i\frac{\omega_L}{v}x}\kappa\int_0^t dt'\tilde{c}(t')\delta(t-x/c-t').
\label{A2}
\end{eqnarray}
Here $\hat{E}_0(x,t)$ is the vacuum field from Eq. (\ref{E0}) with the exponentials $e^{ikx}$ in the mode expansion $\hat{b}_k(0)$. For $x=0^-<0$, the Dirac delta function does not contribute and we indeed obtain the input field $\hat{E}_0(t)-i\Omega_L $. For $x=0^+>0$ we obtain
\begin{eqnarray}
\hat{E}(t)\equiv \hat{E}_0(0^+,t)=\hat{E}_0(t)-i\Omega_L-\kappa\tilde{c}(t).
\label{A3}
\end{eqnarray}
Finally, inserting the coarse-grained solution for $\tilde{c}(t)$ [obtained for simplicity by setting $\dot{\tilde{c}}=0$ in Eq. (\ref{cEOM})], we arrive at Eqs. (\ref{E1}) and (\ref{E3}).

\begin{acknowledgments}
We acknowledge financial support from the Israel Science Foundation (ISF) grant No. 2258/20, the ISF and the Directorate for Defense Research and Development (DDR\&D) grant No. 3491/21, the Center for New Scientists at the Weizmann Institute of Science, the Council for Higher Education (Israel), and QUANTERA (PACE-IN).
This research is made possible in part by the historic generosity of the Harold Perlman Family.
\end{acknowledgments}

\end{document}